\documentclass[12pt]{iopart}
\usepackage{graphicx}
\usepackage{dcolumn}
\usepackage{bm}
\usepackage{color}
\usepackage{hhline}

\begin{document}

\title[STM imaging of HOPG: The role of STM-tip orientations]{Contrast stability and "stripe" formation in Scanning Tunnelling Microscopy imaging of highly oriented pyrolytic graphite: The role of STM-tip orientations}

\author{G\'abor M\'andi$^{1}$, Gilberto Teobaldi$^{2}$, and Kriszti\'an Palot\'as$^{1,3}$}
\address{$^{1}$Budapest University of Technology and Economics, Department of Theoretical Physics,
Budafoki \'ut 8., H-1111 Budapest, Hungary\\
$^{2}$Stephenson Institute for Renewable Energy and Surface Science Research Centre, Department of Chemistry,
University of Liverpool, L69 3BX Liverpool, United Kingdom\\
$^{3}$Condensed Matter Research Group of the Hungarian Academy of Sciences,
Budafoki \'ut 8., H-1111 Budapest, Hungary}
\ead{palotas@phy.bme.hu}

\begin{abstract}

Highly oriented pyrolytic graphite (HOPG) is an important substrate in many technological applications and is routinely used as a
standard in Scanning Tunnelling Microscopy (STM) calibration, which makes the accurate interpretation of the HOPG STM contrast of
great fundamental and applicative importance. We demonstrate by STM simulations based on electronic structure obtained from first
principles that the relative local orientation of the STM-tip apex with respect to the HOPG substrate has a considerable effect on
the HOPG STM contrast. Importantly for experimental STM analysis of HOPG, the simulations indicate that local tip-rotations
maintaining a major contribution of the $d_{3z^2-r^2}$ tip-apex state to the STM current affect only the secondary features of the
HOPG STM contrast resulting in "stripe" formation and leaving the primary contrast unaltered. Conversely, tip-rotations leading to
enhanced contributions from $m\ne 0$ tip-apex electronic states can cause a triangular-hexagonal change in the primary contrast.
We also report a comparison of two STM simulation models with experiments in terms of bias-voltage-dependent STM topography
brightness correlations, and discuss our findings for the HOPG(0001) surface in combination with tungsten tip models of different
sharpnesses and terminations.

\end{abstract}


\section{Introduction}

More than thirty years after the invention of the scanning tunnelling microscope, STM is still one of the most useful tools for
obtaining atomic resolution in surface imaging. However, in spite of such a long and successful history, the interpretation of STM
experiments still raises some, to date unanswered, questions. The explanation of experimental results relies on long-established
electron tunnelling models, which, however, invariably present some level of approximations.
The first tunnelling model presented by Bardeen \cite{bardeen61} was based on first-order perturbation theory.
Tersoff and Hamann \cite{tersoff83,tersoff85} derived a simplified model, where the tip is modelled by a spherically
symmetric wave-function, and the electronic structure of the tip is neglected. Despite its simplicity, the method has successfully
been used for the simulation of STM, and it is still the most commonly used model. However, as was pointed out by Chen
\cite{chen90,chen92}, the symmetry of the tip can have a huge effect on the STM image since the tunnelling matrix
elements are proportional to the derivatives of the sample wave-function depending on the tip orbital symmetries. Tip orbitals
with non-zero orbital momentum (e.g. $p_z$, $d_{3z^2-r^2}$) can lead to an enhancement of the corrugation
\cite{chen90,heinze98}. Later, the roles of the tip orbital symmetry and electronic structure were emphasized in the STM imaging
in several other studies \cite{sacks00,mingo96}. Recently, Palot\'as {\it {et al.}} developed an orbital-dependent tunnelling model
and demonstrated the effect of the tip orbitals on the bias-voltage- and tip-sample distance dependence of the atomic contrast
inversion on the W(110) surface \cite{palotas12orb} and on the Fe(110) surface \cite{mandi14fe}. Extending this model to include
arbitrary tip orientations, they found that different tip orientations can considerably distort the STM image
\cite{mandi13tiprot}. Accordingly, it was suggested that a sound interpretation of experimental STM images cannot,
in principle, be obtained without explicitly accounting for tip-orientation effects.

Recent interest in different carbon allotropes (fullerenes, nanotubes, graphene, graphite) and nanostructures \cite{kanasaki09},
and their potential for a wide spectrum of technological applications \cite{novoselov12,jariwala13,maiti14}, for example
biological and chemical sensors \cite{liu12,pandey14}, nano- and molecular electronics \cite{tapaszto08,otsuki10},
photovoltaics \cite{ayissi13} and catalysis \cite{ruben06,hovel06}, make atomically resolved investigation of carbon substrates
$-$ such as highly oriented pyrolytic graphite (HOPG) $-$ of great relevance across many different scientific fields.

HOPG(0001) is one of the most frequently probed surface, where the tip orbital symmetries play a crucial role.
The tip-dependent corrugation was discussed by Tersoff and Lang,
and the role of the orbital composition of the tip atom was highlighted \cite{tersoff90}. The two nonequivalent carbon atomic
sites of HOPG ($\alpha$ and $\beta$) are responsible for different patterns in STM images. Depending on the applied bias
voltage and tunnelling current both triangular and hexagonal honeycomb patterns can be observed. The selective imaging of the
$\alpha$ and $\beta$ atoms results in a triangular pattern \cite{park86,binnig86}, which is mostly observed under typical
tunnelling conditions, although a honeycomb pattern can be recorded as well \cite{paredes01,wang06}. Chaika {\it {et al.}} showed
that using a [001]-oriented tungsten tip allows for the control of the tip orbitals responsible for the imaging, hence
different patterns in the STM image can be obtained \cite{chaika10,chaika13}. Ondr\'a\v{c}ek {\it {et al.}} showed that
multiple scattering effects can also play an important role in the near contact regime, and they can result in a triangular
pattern in the STM image with hollow sites appearing as bright spots, instead of the carbon atoms \cite{ondracek11}.
Teobaldi {\it {et al.}} rationalised the bias dependent STM contrast mechanisms observed on the HOPG(0001) surface by modelling
a set of tungsten tips taking the effects of tip electronic structure, termination, composition, and sharpness into account
\cite{teobaldi12}.

It is clear that the tip geometry and electronic structure cannot be neglected in an accurate STM simulation method. If the
symmetry of the tip orbitals has a considerable effect on the STM image, it follows naturally that so does the tip orientation.
All simulation methods require a well-defined tip geometry and orientation. Usually a simple geometry is chosen, e.g., a
pyramid-shaped tip apex, but the local tip geometry at the apex and the relative orientation of the sample surface and the tip
apex are unknown and hardly controllable in experiments. Moreover, these tip apex characteristics can even change during the
experimental STM scan, see e.g.\ Refs.\ \cite{hofer08tipH,wasniowska10} for magnetic surfaces.
In separate electronic structure calculations of the sample surface and the tip their local coordinate systems are usually set up
in such a way that they represent the corresponding crystallographic symmetries. The electronic structure data, either the single
electron wave-functions or the density of states (DOS), are defined in the given local coordinate systems, and they are used in
the STM simulations. Thus, the relative orientation of the tip and the sample is fixed, and it usually corresponds to a very
symmetrical setup, which is unlikely in experiments. Hagelaar {\it {et al.}} studied a wide range of tip geometries and
spatial orientations in the imaging of the NO adsorption on Rh(111) in combination with STM experiments \cite{hagelaar08},
and their analysis is quite unique among the published STM simulations.

\begin{figure*}
\includegraphics[width=0.64\textwidth,angle=0]{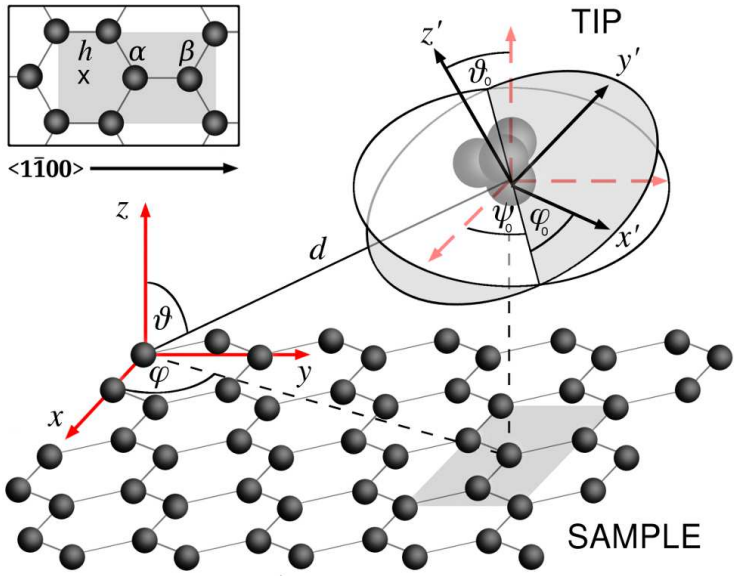}
\caption{\label{Fig1} Schematic view of the STM tip above the HOPG surface. The rotation of the tip local
coordinate system with respect to that of the sample surface is described by the Euler angles $(\vartheta_0,\varphi_0,\psi_0)$.
The shaded rectangle shows the considered scanning area of the HOPG surface for STM simulations in Figs.\ \ref{Fig5}, \ref{Fig6},
and \ref{Fig7}. The positions of the characteristic $h$, $\alpha$, and $\beta$ sites of the HOPG(0001) surface are indicated in
the inset.
}
\end{figure*}

In the present work we employ the three-dimensional (3D) Wentzel-Kramers-Brillouin (WKB) electron tunnelling theory implemented in
the 3D-WKB-STM code \cite{palotas12orb,mandi14fe,mandi13tiprot,palotas11sts,palotas11stm,palotas12sts,palotas13fop} to study the
STM contrast characteristics of the HOPG(0001) surface as a function of the local orientation of a set of tungsten tips.
In the tunnelling model the tip orientation, defined by the local coordinate system of a crystallographically well-defined tip
surface with ($hkl$) Miller indices, can be rotated by the Euler angles $\left(\vartheta_{0},\varphi_{0},\psi_{0}\right)$ in an
arbitrary fashion \cite{mandi13tiprot}. A schematic view of an STM tip with rotated local coordinate system above the HOPG(0001)
surface is shown in Fig.\ \ref{Fig1}. The systematic effect of local tip rotations is practically unexplored in experiments.
The reason is that there is no direct in-situ information about the local rearrangements of the tip apex
structure, e.g., manifesting as local tip rotations during scanning in an STM equipment, and the atomically precise stability
of the tip apex structure is almost impossible to control.
As we will demonstrate theoretically, local tip rotations can have an important effect on the STM contrast.
Initially, we compare the 3D-WKB and Bardeen tunnelling models with each other and with experimental results using
bias dependent topography brightness correlations. We find quantitatively good agreement for particular tips and bias voltage
ranges, and discuss the identified differences. Based on the comparison with experimental data we conclude that the
two tunnelling methods perform at the same quantitative reliability at both positive and negative bias voltages.

The paper is organised as follows: After a brief description of computational details in Sec.\ \ref{sec_comp}, we define the
topography brightness, and compare the 3D-WKB method with Bardeen's approach in terms of correlations between the calculated
relative brightnesses above the HOPG surface in section \ref{sec_res_compar_Bardeen}. Comparison with available experimental data
\cite{teobaldi12} is reported in section \ref{sec_res_compar_exp}. The simulated effect of the local tip orientation on the
STM image contrast is presented in section \ref{sec_res_stm}, followed by our conclusions in section \ref{sec_conc}.
The 3D-WKB tunnelling theory with an arbitrary tip orientation is briefly presented in Appendix.

\section{Computational details}
\label{sec_comp}

The HOPG(0001) surface and a set of tungsten tips were modelled in Ref.\ \cite{teobaldi12}. Slab geometry relaxations were
performed and the PDOS of the tip apex and sample surface atoms were calculated within the generalised gradient approximation
Perdew-Burke-Ernzerhof (GGA-PBE) \cite{pbe96} projector augmented wave (PAW) scheme implemented in the plane-wave VASP code
\cite{VASP2,VASP3,kresse99}. Details on the geometries of the HOPG surface and the W tips as well as on the performed electronic
structure calculations are found in Fig.\ 1 and Sec. II B of Ref.\ \cite{teobaldi12}.

For the 3D-WKB STM simulations we chose $\phi_S=\phi_T=4.8$ eV electron work function for both the HOPG surface \cite{shiraishi01}
and the tungsten tips \cite{palotas12orb}. The tunnelling current was calculated in a box above the rectangular scan area of the
HOPG(0001) surface shown as the shaded area in the inset of Fig.\ \ref{Fig1} containing $31\times 21$ lateral grid points in
accordance with the STM calculations of Ref.\ \cite{teobaldi12} using the Bardeen approach. This corresponds to 0.142 \AA$\;$and
0.123 \AA$\;$resolution in the $x$ and $y$ direction, respectively, and in the surface-normal $z$ direction we used a finer,
0.02 \AA$\;$resolution. The constant-current contours are extracted following the method described in Ref.\ \cite{palotas11stm},
and we report STM images above the mentioned rectangular scan area.
In Eq.(\ref{Eq_current}) the atomic superposition (sum over $i$) has to be carried out, in principle, over all surface atoms.
Convergence tests, however, show that taking a relatively small number of atoms into account provides converged current values
because of the exponentially decaying electron states into the vacuum \cite{palotas12orb,palotas11sts}. We also found that the
tip orientation and geometry do not affect this convergence significantly \cite{mandi13tiprot}. In the case of calculating
STM images of the HOPG surface, we considered carbon atoms which are at most $d_{||}=7.5$ \AA$\;$far from the edge of the
scan area, thus involving altogether 117 surface atoms in the atomic superposition.

Employing the BSKAN code \cite{hofer03pssci,palotas05} it was pointed out in Ref.\ \cite{teobaldi12}
that the tunnelling current depends on the relative orientation of the tip and the surface, and two orthogonal orientations were
considered for three tip models with different sharpnesses and compositions: $\mathrm{(r)W_{blunt}}$, $\mathrm{(r)W_{sharp}}$,
and $\mathrm{(r)W_{C-apex}}$, with "r" marking the tips rotated by 90 degrees around the $z$ axis normal to the surface plane.
In the 3D-WKB model an arbitrary tip rotation can be
performed by setting the corresponding Euler angles $\left(\vartheta_{0},\varphi_{0},\psi_{0}\right)$, see also Fig.\ \ref{Fig1}.
Due to our choice of the fixed sample and tip geometries the rotated (rW) tips of Ref.\ \cite{teobaldi12} correspond to
$(0^{\circ},0^{\circ},0^{\circ})$, and the unrotated (W) tips to $(0^{\circ},0^{\circ},90^{\circ})$ Euler angles.
Note that when changing the Euler angles, tunnelling through one tip apex atom was considered only, and contributions
from other tip atoms were not taken into account. High degrees of tilting the tip ($\vartheta_0>30^{\circ}$) could, in fact,
result in multiple tip apices \cite{rodary11} depending on the local geometry, which can increase the tunnelling current,
but can also lead to the destruction of the atomic resolution in STM images.

\section{Results and discussion}
\label{sec_res}

To demonstrate the reliability of the 3D-WKB approach, first we perform a systematic comparison of bias-dependent normalised
constant-current topographs (relative brightnesses) calculated above the HOPG surface with those obtained by Bardeen's
tunnelling approach. We discuss the differences and their origins. Comparing the simulated relative brightnesses with
experimental data \cite{teobaldi12} we find that the two tunnelling methods perform at the same quantitative reliability.
Turning to STM images, we show that the local tip orientation has a considerable effect on the obtained constant-current contrast.

For the analysis of the topographic contrast we calculate brightness profiles along the $\langle 1\bar{1}00\rangle$ direction
of the HOPG(0001) surface, following the methods described in Ref.\ \cite{teobaldi12}. These brightness profiles are line sections
of the constant-current contour at a given bias voltage, which contain the three characteristic positions of the HOPG surface:
hollow ($h$), carbon-$\alpha$, and carbon-$\beta$, see inset of Fig.\ \ref{Fig1}. In order to compare the brightness profiles
of different tip geometries and bias voltages, the profiles are scaled to the [0,1] interval. The definition of the relative
brightness of a given point ($\mathbf{x}$) along the scan line is the following:
\begin{equation}
\label{Eq_brightness}
B(\mathbf{x},V)=\frac{z(\mathbf{x},V)-z(\mathbf{x}_{min},V)}{z(\mathbf{x}_{max},V)-z(\mathbf{x}_{min},V)},
\end{equation}
where $z(\mathbf{x},V)$ is the height of the constant-current contour above the $\mathbf{x}$ point at bias voltage $V$,
$z(\mathbf{x}_{min},V)$ and $z(\mathbf{x}_{max},V)$ respectively have the smallest and largest apparent heights along the scan
line, thus $B(\mathbf{x}_{min},V)=0$ and $B(\mathbf{x}_{max},V)=1$. The current values were chosen for each bias voltage
in the interval of [-1 V, 1 V] in steps of 0.1 V in such a way that the lowest apparent height of each constant-current contour
was 5.5 \AA.

Using the same lateral resolution of the scanning area employing two different methods $M1$ and $M2$, it is possible to
quantitatively compare the relative brightness profiles $B_{M1}$ and $B_{M2}$ by calculating the correlation coefficient as
\begin{eqnarray}
&&r_{B_{M1}B_{M2}}(V)=\nonumber\\
&&\frac{\sum_{k=1}^{n}[B_{M1}(x_k,V)-\overline{B}_{M1}(V)][B_{M2}(x_k,V)-\overline{B}_{M2}(V)]}{\sqrt{\left(\sum_{k=1}^{n}[B_{M1}(x_k,V)-\overline{B}_{M1}(V)]^2\right)\left(\sum_{k=1}^{n}[B_{M2}(x_k,V)-\overline{B}_{M2}(V)]^2\right)}}.
\label{Eq_correlation}
\end{eqnarray}
Here, $\overline{B}_{Mi}(V)=\frac{1}{n}\sum_{k=1}^{n}B_{Mi}(x_{k},V)$ is the mean value of the brightness profile obtained by
method $Mi$ at bias voltage $V$, and $B_{Mi}(x_{k},V)$ denotes the relative brightness of the $k$th point of the $B_{Mi}$
profile, which consists of $n$ points. In this paper we compare the following:
$Mi\in$ \{3D-WKB, Bardeen, Experiment\}, the data for the last two were taken from Ref.\ \cite{teobaldi12}.

\subsection{Comparison between 3D-WKB and Bardeen methods}
\label{sec_res_compar_Bardeen}

\begin{figure*}
\includegraphics[width=1.08\textwidth,angle=0]{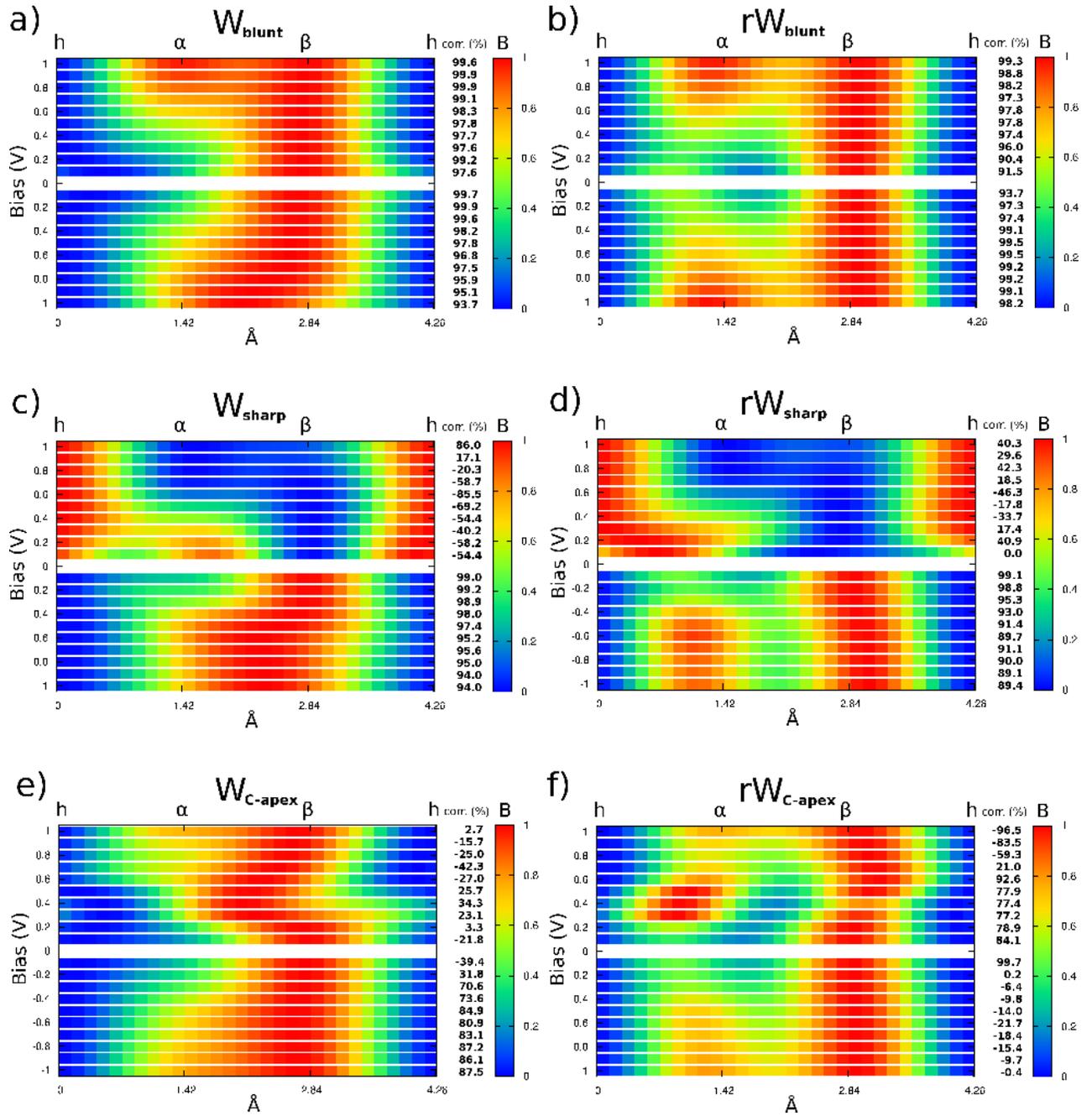}
\caption{\label{Fig2} Relative brightness profiles $B(\mathbf{x},V)$ in Eq.(\ref{Eq_brightness}) along the
$\langle 1\overline{1}00\rangle$ direction ($h-\alpha -\beta -h$ line) of the HOPG(0001) surface calculated by the 3D-WKB method,
and percentual correlations following Eq.(\ref{Eq_correlation}) with those obtained by the Bardeen approach at given
bias voltages in the [-1 V, 1 V] range for different tip models: a) $\mathrm{W_{blunt}}$, b) $\mathrm{rW_{blunt}}$,
c) $\mathrm{W_{sharp}}$, d) $\mathrm{rW_{sharp}}$, e) $\mathrm{W_{C-apex}}$, f) $\mathrm{rW_{C-apex}}$, see text for details.
}
\end{figure*}

Using the correlation coefficient defined in Eq.(\ref{Eq_correlation}), we compare the relative brightness profiles obtained by
the 3D-WKB and Bardeen methods. Fig.\ \ref{Fig2} shows bias-dependent relative brightnesses above the $h-\alpha -\beta -h$ line
along the $\langle 1\bar{1}00\rangle$ direction of the HOPG(0001) surface in the bias voltage range of [-1 V, 1 V] in steps of 0.1
V for each considered W and rW tip models using the 3D-WKB method. The corresponding relative brightness profiles obtained by the
Bardeen method can be found in Fig.\ 9 of Ref.\ \cite{teobaldi12}. Fig.\ \ref{Fig2} also presents the calculated percentual
correlations between the brightness profiles of the two methods at each bias voltage.

We also calculate correlations considering the negative (-1 V$\leq V<$ 0 V), positive (0 V$<V\leq$ 1 V), and full
(-1 V$\leq V\leq$ 1 V) bias ranges. In these cases the $B_{\mathrm{3D-WKB}}(x_k,V)$
and $B_{\mathrm{Bardeen}}(x_k,V)$ brightness data consist of ten (negative or positive bias) or twenty (full bias range) times the
number of points $(n=31)$ of a single bias brightness profile. The results are listed in Table \ref{Tab1}.

\begin{table}[ht]
\begin{centering}
\begin{tabular}{|c|c|c|c|c|c|c|}
\hline
\textbf{3D-WKB vs.\ Bardeen} & $\mathrm{W_{blunt}}$ & $\mathrm{rW_{blunt}}$ & $\mathrm{W_{sharp}}$ & $\mathrm{rW_{sharp}}$ & $\mathrm{W_{C-apex}}$ & $\mathrm{rW_{C-apex}}$ \tabularnewline\hline
negative bias   & 97.1 & 98.3 & 96.1 & 92.4 & 62.0 &  3.4 \tabularnewline\hline
positive bias   & 98.3 & 96.5 &-27.1 &  8.9 & -4.9 & 27.3 \tabularnewline\hline
full bias range & 97.5 & 97.3 & 36.6 & 48.8 & 29.2 & 16.0 \tabularnewline\hline
\end{tabular}
\end{centering}
\caption{Percentual relative brightness correlations according to Eq.(\ref{Eq_correlation}) between the 3D-WKB and Bardeen
methods for different tip models in the negative (-1 V$\leq V<$ 0 V), positive (0 V$<V\leq$ 1 V), and full (-1 V$\leq V\leq$ 1 V)
bias ranges.}
\label{Tab1}
\end{table}

Considering the obtained correlations, we find an excellent agreement between the 3D-WKB and the Bardeen brightness results in the
case of the $\mathrm{W_{blunt}}$ and $\mathrm{rW_{blunt}}$ tips [Figs.\ \ref{Fig2}a) and \ref{Fig2}b)]. All of the single bias
profiles show at least 90\% correlation, and in the full bias range the correlation is more than 97\% for both orientations. For
the $\mathrm{W_{sharp}}$ and $\mathrm{rW_{sharp}}$ tips [Figs.\ \ref{Fig2}c) and \ref{Fig2}d)] a good agreement between the two
models is found at negative bias voltages only, where the brightness profiles are qualitatively similar to the ones obtained by
the blunt tip models. In the positive bias range the 3D-WKB model shows that the $h$ position has the largest apparent height at
almost each considered bias voltage, and in effect, the STM contrast is reversed at positive compared to negative bias voltages.
We return to this asymmetry later on. For the $\mathrm{W_{C-apex}}$ and $\mathrm{rW_{C-apex}}$ tips [Figs.\ \ref{Fig2}e) and
\ref{Fig2}f)] the agreement is the poorest between the two tunnelling models.

These results can be rationalised on the basis of the different contributions of the orbital-decomposed tip electronic states to
the tunnelling current, and can be explained by the atomic geometry of the STM tip models in view of the different concepts of the
tunnelling models. The Bardeen method uses the Kohn-Sham single electron states in the vacuum to construct the transmission matrix
elements, i.e., outside the localisation radii of the PAW projectors. On the other hand, in the 3D-WKB model it is assumed that
electrons tunnel through one tip apex atom, and the PDOS of this apex atom is used for describing the tip electronic structure
which is constructed based on the PAW projectors. The exponential decay of the electron states into the vacuum is taken into
account by the transmission coefficient in Eq.(\ref{Eq_Transmission}). The PDOS of the tip apex atom is sensitive to the chemical
environment, i.e., to the quality and geometrical arrangement of the surrounding atoms. In case of the $\mathrm{(r)W_{blunt}}$
tips the PDOS of the tip apex represents well the electronic structure of the whole tip, and there is practically no significant
difference in the description of the tunnelling process between the two methods. For the $\mathrm{(r)W_{sharp}}$ and
$\mathrm{(r)W_{C-apex}}$ tips a pyramidal atomic arrangement was considered, and the transmission functions differ considerably
in the two methods. For example, in case of the $\mathrm{(r)W_{C-apex}}$ tips the W atoms from the pyramid itself are expected to
contribute much more to the tunnelling due to their relatively large $d$-DOS compared to the C-apex $p$-DOS, see Fig.\ 6 of
Ref.\ \cite{teobaldi12}. These electron states are considered in the Bardeen but not in the 3D-WKB model.

To understand the practically reversed brightness profiles at positive with respect to negative bias voltages for the
$\mathrm{(r)W_{sharp}}$ tips a deeper analysis is needed. As a first indication, it was found that the local density of states
(LDOS) 3 \AA$\;$above the tip apex is much more asymmetric in the bias voltage for the $\mathrm{W_{sharp}}$ than for the
$\mathrm{W_{blunt}}$ tip, see Fig.\ 6(d) of Ref.\ \cite{teobaldi12}. The 3D-WKB method allows for the decomposition of the
tunnelling current according to the orbital symmetries $\sigma$ (sample) and $\tau$ (tip): $I_{\sigma\tau}$.
The electronic structure calculation of the HOPG sample showed that the $p_z$-like PDOS is
at least an order of magnitude larger than the $s$-, $p_x$- and $p_y$-like PDOS for both $\alpha$- and $\beta$-type carbon atoms
in the range of $\pm$1 eV around the Fermi energy. This means that the HOPG electronic structure can safely be approximated by
taking the $p_z$-like PDOS only, and we fixed the orbital index of the sample as $\sigma=p_z$. On the other hand, the W-apex has
$\tau\in\{s,p_y,p_z,p_x,d_{xy},d_{yz},d_{3z^2-r^2},d_{xz},d_{x^2-y^2}\}$, and the C-apex has $\tau\in\{s,p_y,p_z,p_x\}$ orbital
symmetries in the considered tip models.

\begin{figure*}[h]
\includegraphics[width=0.67\textwidth,angle=0]{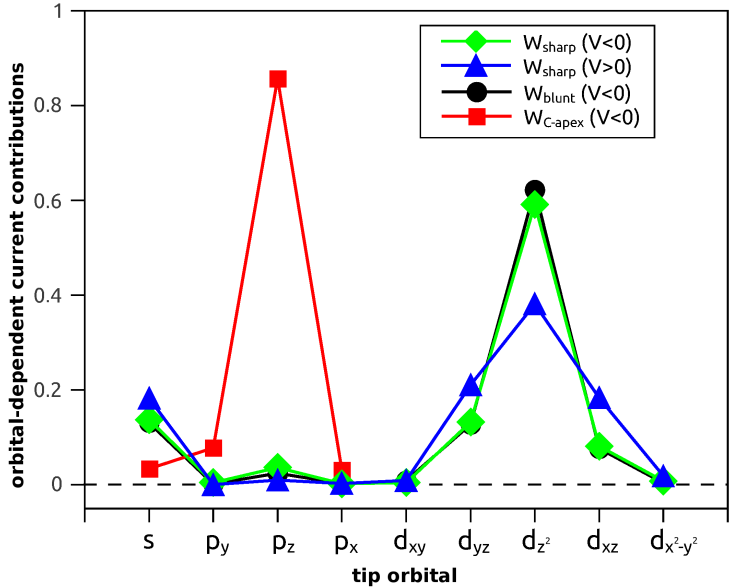}
\caption{\label{Fig3} Orbital-dependent relative current contributions $\tilde{I}_{\sigma\tau}$ defined in
Eq.(\ref{Eq_i_bg}) for $\sigma=p_z$, 5.5 \AA$\;$above the $\beta$ atom of the HOPG(0001) surface at $\pm$1 V bias
voltages using three different tip models. The tip orbitals ($\tau$) are explicitly shown.
For brevity, we used the notation of $d_{z^2}$ for the $d_{3z^2-r^2}$ orbital.
}
\end{figure*}

Using Eq.(\ref{Eq_i_bg}) we calculate the relative contribution of all $\sigma\leftrightarrow\tau$ transitions to the tunnelling
current, $\tilde{I}_{\sigma\tau}$, 5.5 \AA$\;$above the $\beta$ carbon atom at $\pm$1 V bias voltages. The current histograms shown
in Fig.\ \ref{Fig3} give the percentual contributions of the different tip orbitals to the current for the three tip models.
First, let us focus on the bias-asymmetry of the contributions of the $\mathrm{W_{sharp}}$ tip. For this case the $s$, $d_{yz}$,
$d_{3z^2-r^2}$, and $d_{xz}$ tip states are dominant, and the largest contribution comes from the $d_{3z^2-r^2}$ state.
As can clearly be seen, the main difference in the positive and negative bias ranges shows up in the increasing $d_{yz}$ and
$d_{xz}$ contributions with a concomitant decreasing of the $d_{3z^2-r^2}$ contribution at positive bias. Since $m\ne 0$ tip
states are responsible for a contrast inversion on metal surfaces \cite{chen92,palotas12orb}, these current histograms explain
the observed contrast inversion with respect to the bias polarity above the $\beta$ carbon atom of the HOPG surface found in
Figs.\ \ref{Fig2}c) and \ref{Fig2}d). Based on the current histograms, we also expect that the $\mathrm{W_{sharp}}$ and
$\mathrm{W_{blunt}}$ tips provide similar contrast at negative bias voltages. This is confirmed by Fig.\ \ref{Fig2}.
Note that changing the bias voltage in the respective negative (-1 V$\leq V<$ 0 V) and positive (0 V$<V\leq$ 1 V) ranges does not
influence the quality of the current histograms. For the $\mathrm{W_{blunt}}$ and $\mathrm{W_{C-apex}}$ tips no qualitative
difference of the current histograms were found at positive bias voltages, therefore, the $V<$ 0 V results are shown only.
Moreover, it is seen in Fig.\ \ref{Fig3} that the largest contribution is due to the $p_z-p_z$ transition for the
$\mathrm{W_{C-apex}}$ tip: it gives 85\% of the total current.

These features of the current histograms can be understood from the energy dependence of the PDOS of the tip apices, and also from
the angular dependence of the electron states. In Fig.\ 6 of Ref.\ \cite{teobaldi12} one can see that for the
$\mathrm{W_{blunt}}$ and $\mathrm{W_{C-apex}}$ tips the PDOS functions are fairly symmetric with respect to the Fermi
energy, thus the bias voltage does not affect the current contributions significantly. Although some of the orbitals have rather
asymmetric PDOS, these give small contributions to the tunnelling current due to their angular dependence, thus they do
not affect the histograms, e.g., the $p_x$ state of the $\mathrm{W_{C-apex}}$ tip, or the $d_{x^2-y^2}$ state of the
$\mathrm{W_{blunt}}$ tip. On the other hand, the PDOS functions of the $\mathrm{W_{sharp}}$ tip apex are rather asymmetric,
particularly for the $d_{3z^2-r^2}$ state, which has the largest contribution. For $E>E_F^T$, which is relevant at negative bias,
it is larger than for $E<E_F^T$, thus the current contribution is also larger for negative bias, as seen in Fig.\ \ref{Fig3}.
All in all, this asymmetric behaviour of the PDOS of the $\mathrm{W_{sharp}}$ tip apex is responsible for the observed
contrast inversion with respect to the bias polarity in Figs.\ \ref{Fig2}c) and \ref{Fig2}d).

\subsection{Comparison between simulations and experiment}
\label{sec_res_compar_exp}

In experimental STM images of HOPG, it is possible to identify the $\langle 1\bar{1}00\rangle$ direction (assuming that the
brightest features lie along this direction), however, the order of $h$, $\alpha$, and $\beta$ sites is unknown,
see Figs.\ 3 and 4 of Ref.\ \cite{teobaldi12}. The only possible way to determine the $h-\alpha -\beta$ or $h-\beta -\alpha$ order
along the $\langle 1\bar{1}00\rangle$ direction is the direct comparison of experimental and simulated brightness profiles.
Since the experimental profiles are obtained by averaging numerous sections of the scan lines (for more information, see
Ref.\ \cite{teobaldi12}), the comparison at different bias voltages can be performed if the profiles are transformed to start with
their corresponding maximum or minimum.
While in Ref.\ \cite{teobaldi12} the relative brightness profiles are shifted to start with their maximum, we transform them to
start with their global minimum. The motivation for changing the reference point is the following: the experimental
brightness profiles at each bias voltage have one minimum only, while at certain voltages they have two local maxima very close in
magnitude to each other: $B(x_{max_{1}},V)\approx B(x_{max_{2}},V)$, similarly to the simulated brightness profiles using the
$\mathrm{rW_{blunt}}$ tip at larger bias voltages, see Fig.\ \ref{Fig2}b). If the profiles are shifted to start
with the global maximum then the correlation coefficient strongly depends on the actual position ($x_{max_{1}}$ or $x_{max_{2}}$)
of the global maximum. For example, when comparing two almost identical brightnesses with two local maxima at $\alpha$ and $\beta$
sites, if the global maximum in one profile is $\alpha$, and is $\beta$ in the other, then the correlation coefficient
of the two profiles shifted to the corresponding global maximum can be negative, instead of the value of close to 1.
Rigidly shifting the brightness profiles to start with their global minimum value solves this problem.

\begin{figure*}[h]
\includegraphics[width=1.08\textwidth,angle=0]{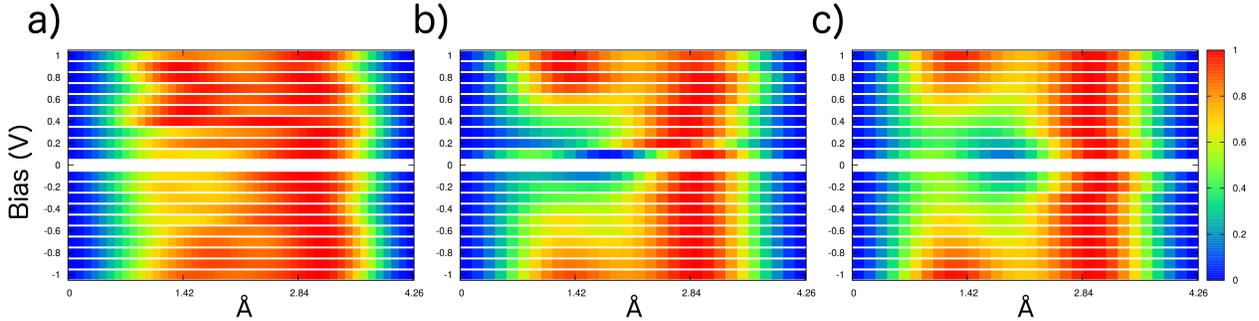}
\caption{\label{Fig4} Experimental and simulated relative brightness profiles $B(\mathbf{x},V)$ in
Eq.(\ref{Eq_brightness}) along the $\langle 1\overline{1}00\rangle$ direction ($h-\alpha -\beta -h$ line) of the HOPG(0001)
surface: a) Experiment \cite{teobaldi12}, b) Bardeen, c) 3D-WKB. All profiles are rigidly shifted to start with their
global minimum value. In the simulations the $\mathrm{rW_{blunt}}$ tip was used.
}
\end{figure*}

Following this convention, Fig.\ \ref{Fig4} shows a comparison between the experimental \cite{teobaldi12},
Bardeen-calculated and 3D-WKB modelled brightness profiles. In the simulations the $\mathrm{rW_{blunt}}$ tip was used.
We obtain good qualitative agreement on the bias-dependence of the triangular-hexagonal transition between the experiments and
simulations. To quantify the agreement the correlation coefficients between the experimental and simulated brightness profiles are
reported in Table \ref{Tab2} using all of the previously introduced tip models.

\begin{table}[ht]
\begin{centering}
\begin{tabular}{|c|c|c|c|c|c|c|}
\hline
\textbf{Bardeen vs.\ Experiment} & $\mathrm{W_{blunt}}$ & $\mathrm{rW_{blunt}}$ & $\mathrm{W_{sharp}}$ & $\mathrm{rW_{sharp}}$ & $\mathrm{W_{C-apex}}$ & $\mathrm{rW_{C-apex}}$ \tabularnewline\hline
negative bias   & 91.3 & 92.6 & 89.8 & 84.6 & 93.5 & 91.9 \tabularnewline\hline
positive bias   & 90.6 & 88.2 & 67.2 & 69.8 & 87.8 & 78.8 \tabularnewline\hline
full bias range & 90.9 & 90.2 & 77.6 & 75.0 & 90.5 & 85.1 \tabularnewline\hline\hline
\textbf{3D-WKB vs.\ Experiment}  & $\mathrm{W_{blunt}}$ & $\mathrm{rW_{blunt}}$ & $\mathrm{W_{sharp}}$ & $\mathrm{rW_{sharp}}$ & $\mathrm{W_{C-apex}}$ & $\mathrm{rW_{C-apex}}$ \tabularnewline\hline
negative bias   & 90.7 & 92.5 & 89.9 & 85.3 & 93.6 & 91.9 \tabularnewline\hline
positive bias   & 89.9 & 87.9 & 66.1 & 68.1 & 87.4 & 78.5 \tabularnewline\hline
full bias range & 90.3 & 90.0 & 77.0 & 74.3 & 90.3 & 84.9 \tabularnewline\hline
\end{tabular}
\end{centering}
\caption{Percentual relative brightness correlations according to Eq.(\ref{Eq_correlation}) between the simulated results
(Bardeen, 3D-WKB) using different tip models and the experimental data (see Fig.\ \ref{Fig4}a) and
Ref.\ \cite{teobaldi12}) in the negative (-1 V$\leq V<$ 0 V), positive (0 V$<V\leq$ 1 V), and full (-1 V$\leq V\leq$ 1 V) bias
ranges.}
\label{Tab2}
\end{table}

The two tunnelling methods produce almost the same correlation coefficients when comparing the simulated brightness
profiles with the experimental results, the difference between them is always less then 2\%. This finding is independent of the
applied tip model or bias polarity. Based on the correlation values, we also find that brightness profiles of the
$\mathrm{(r)W_{blunt}}$ and $\mathrm{(r)W_{C-apex}}$ tips are very similar to the experimental ones, while the
$\mathrm{(r)W_{sharp}}$ tip models perform better at negative compared to positive bias polarity.

\subsection{STM images}
\label{sec_res_stm}

\begin{figure*}[h]
\includegraphics[width=0.85\textwidth,angle=0]{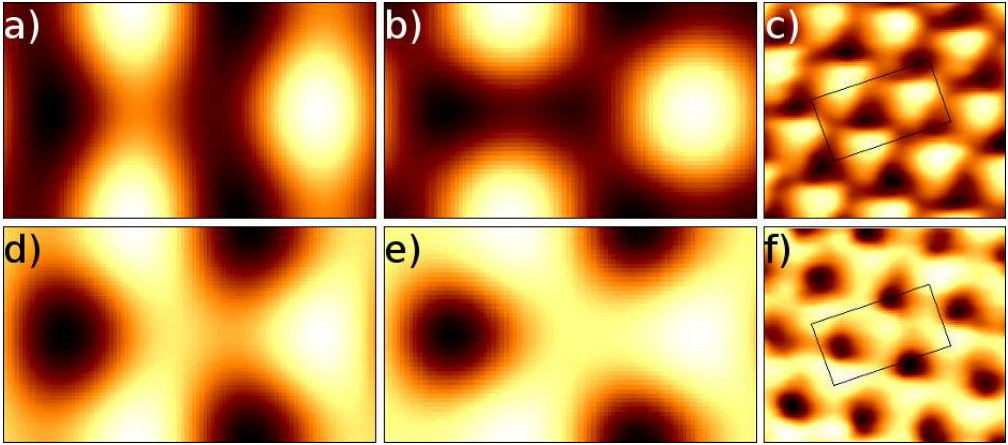}
\caption{\label{Fig5} Bias voltage effect on the simulated STM image contrast of HOPG at a fixed
$(\vartheta_0=0^{\circ},\varphi_0=0^{\circ},\psi_0=0^{\circ})$ tip orientation using the blunt W tip:
first row a)-c) 0.1 V, second row d)-f) 0.6 V bias voltage; a), d) 3D-WKB, b), e) Bardeen. For comparison,
experimental STM images \cite{teobaldi12} are also shown: c) and f), with the rectangular scan area for the simulations (see also
the inset of Fig.\ \ref{Fig1}). The qualities of the STM image contrasts correspond to the results of
Fig.\ \ref{Fig4}.
}
\end{figure*}

\begin{figure*}
\includegraphics[width=1.00\textwidth,angle=0]{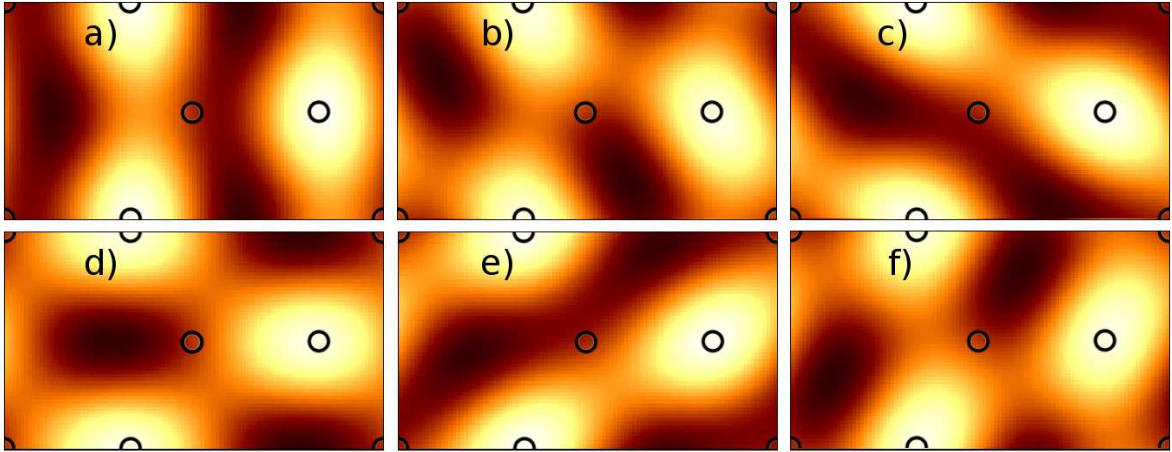}
\caption{\label{Fig6} Tip rotation effect on the simulated STM images of HOPG at $V=0.1$ V using the blunt W tip.
$\vartheta_0=\psi_0$ are fixed at 0 degrees in each part. Parts a)-f) correspond to $\varphi_0$ values of 0, 30, 60, 90, 120,
and 150 degrees, respectively. The atomic positions are denoted by circles, and the rectangular scan area is shown in the inset of
Fig.\ \ref{Fig1}. Note that image Fig.\ \ref{Fig6}a) is the same as Fig.\ \ref{Fig5}a).
}
\end{figure*}

\begin{figure*}
\includegraphics[width=0.67\textwidth,angle=0]{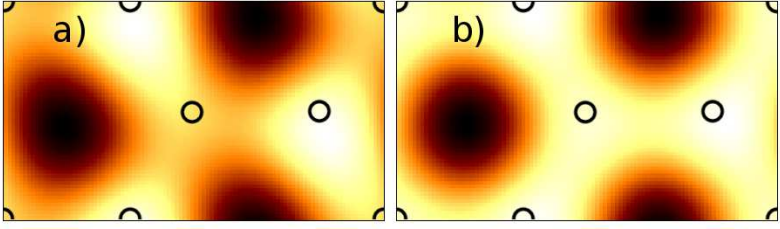}
\caption{\label{Fig7} Tip rotation effect on the simulated STM image contrast of HOPG at $V=0.7$ V using the
blunt W tip. Parts a) and b) correspond to $(\vartheta_0=15^{\circ},\varphi_0=25^{\circ},\psi_0=0^{\circ})$ and
$(\vartheta_0=15^{\circ},\varphi_0=150^{\circ},\psi_0=0^{\circ})$ tip orientations, respectively.
The atomic positions are denoted by circles, and the rectangular scan area is shown in the inset of Fig.\ \ref{Fig1}.
A triangular-hexagonal contrast change is observed due to the tip rotation.
}
\end{figure*}

To investigate the STM contrast changes depending on the bias voltage and on the tip orientation, constant-current STM images are
simulated. The calculated images shown in Figs.\ \ref{Fig5}, \ref{Fig6}, and \ref{Fig7} are taken in the
rectangular scan area shown in the inset of Fig.\ \ref{Fig1}, and all contours have the same minimum apparent height of 5.5 \AA.
We use the convention for the definition of the two different contrast patterns as in Ref.\ \cite{teobaldi12}: A triangular
pattern has two brightness maxima in the scan area, and beside these a hexagonal pattern has two secondary maxima with
relative brightness larger than 0.7.

In Fig.\ \ref{Fig5} we demonstrate the bias-dependent contrast change at two characteristic bias voltages
for both 3D-WKB and Bardeen methods employing a $\mathrm{rW_{blunt}}$ tip,
and compare the simulation results to experiments \cite{teobaldi12}.
From the brightness profiles of Fig.\ \ref{Fig4} we expect a triangular pattern of bright spots for
0.1 V bias voltage as these profiles have one global maximum. On the other hand, a hexagonal honeycomb pattern is
expected for 0.6 V bias as the corresponding profiles have two local maxima.
These expectations are in accordance with the simulated constant-current STM images of Fig.\ \ref{Fig5}a)-b) at
0.1 V and \ref{Fig5}d)-e) at 0.6 V, and we obtain a qualitatively good agreement for the primary contrast in comparison with
experiments shown in Fig.\ \ref{Fig5}c) at 0.1 V and \ref{Fig5}f) at 0.6 V. Thus, the results confirm that the bias voltage has a
major influence on the apparent height of the atoms in the STM images of HOPG \cite{teobaldi12}.

To investigate the effect of the tip orientation on the STM contrast, we simulate constant-current STM images of the HOPG surface
at 0.1 V bias voltage using the $\mathrm{W_{blunt}}$ tip with different local orientations of the apex. First, tip rotations
around the $z=z'$-axis are considered, i.e., we fix the Euler angles $\vartheta_0=\psi_0=0^{\circ}$, and change $\varphi_0$ from
$0^{\circ}$ to $150^{\circ}$ in $30^{\circ}$ steps (see Fig.\ \ref{Fig1}). This way, no orientational change of the dominating
$d_{3z^2-r^2}$ tip-apex orbital state is present \cite{mandi13tiprot}. The obtained constant-current STM images are shown in
Fig.\ \ref{Fig6}. We find that the primary features of the images do not change with such kind of tip rotations: the maxima of the
contours are always located at the same carbon-$\beta$ positions, thus the images preserve the symmetry of the
HOPG surface, and the tip is stable using the experimentalist terminology.
At the selected bias voltage and tip-sample distance we observe
a triangular pattern with the apparent height of the $\beta$ atoms significantly larger than that of the $\alpha$ atoms. The
effect of the tip rotation shows up as a secondary feature in the STM images. There are certain lateral directions where the
apparent heights are larger and elongated, thus we can identify "stripes" in the images. The direction of these "stripes" is
independent of the underlying atomic structure of the HOPG surface, thus it is clearly the rotational effect of
the blunt W(110) tip having $C_{2v}$ symmetry. Note that similar elongated features are also reported in Fig.\ 15(b)
of Ref.\ \cite{tsukada91} for the HOPG surface using a blunt W(110) cluster model for the STM tip.
Similar "stripes" can also be observed in experimental STM images, see e.g.,
Fig.\ 3 of Ref.\ \cite{teobaldi12}. It was even found that the "stripes" can change their lateral orientation depending on the
bias voltage (compare Figs.\ 3(f) and 3(i) of Ref.\ \cite{teobaldi12}). According to our interpretation, this suggests two
differently rotated local tip apex geometries at the two bias voltages.
We note that in-plane low-barrier sub-apex atomic rearrangements, while maintaining the tip-apex, can lead to an effective
rotation of the tip-apex structure (see for instance the models in Figs.\ 1(b) and 1(d) in Ref.\ \cite{teobaldi12}), causing the
simulated and measured changes in the STM "stripes".

On the other hand, it is interesting to find that the primary features of the STM image can change by the same kind of local tip
rotation around the $z'$-axis by $\varphi_0$. The requirement for this is a non-zero $\vartheta_0$, i.e., a tilted $d_{3z^2-r^2}$
tip-apex orbital with respect to the surface normal of the substrate.
Fig.\ \ref{Fig7} demonstrates that the STM image contrast can change between the triangular and hexagonal patterns
above the HOPG surface solely due to the change of the tip orientation by fixing all other tunnelling parameters.
For this case we selected 0.7 V bias voltage, and two orientations of the $\mathrm{W_{blunt}}$ tip:
$(\vartheta_0=15^{\circ},\varphi_0=25^{\circ},\psi_0=0^{\circ})$ and
$(\vartheta_0=15^{\circ},\varphi_0=150^{\circ},\psi_0=0^{\circ})$.
Note that the modelled contrast change is obtained at a $125^{\circ}$ difference in $\varphi_0$, and is expected to be
due to enhanced contributions from $m\ne 0$ tip-apex electronic states to the tunnelling current upon tip-rotation
\cite{mandi13tiprot}. As a further consequence, our simulations indicate that tip instabilities in STM experiments are likely
found for local tip-apex geometries described by non-zero $\vartheta_0$ angles that also result in distorted STM images
\cite{mandi13tiprot}.

Note that tip-surface interactions can further complicate the STM contrast. In Ref.\ \cite{ondracek11} it was
shown that multiple scattering effects can induce a contrast change shifting the maximum brightness from $\beta$ carbon to the
hollow position above the HOPG surface in the near contact regime (below 4 \AA\;of tip-sample separation). Since our minimum
tip-sample distance is always 5.5 \AA, i.e., we are in the pure tunnelling regime, we expect that the tip-surface force is
monotonically decreasing with decreasing current by moving the tip away from the surface. Thus force related changes in the
contrast do not modify our conclusions on the effect of the tip orientations observed in STM images in pure tunnelling regimes
for tip-surface distances larger than 4 \AA. However, close to the contact substantial effects of the tip-surface force on the
STM contrast can be expected upon tip rotation, which could be interesting to study in the future using an appropriate method.

Overall, our findings strongly point to a non-negligible role of the local tip orientation or sub-apex
rearrangements for the STM contrast of HOPG surfaces. They also suggest that the tip-apex orientation may have marked effects on
the STM appearances of other substrates, and this should be accounted for in STM simulations if aiming at accuracy.
In this respect, and given its very favourable computational cost, the 3D-WKB atom-superposition electron tunnelling model
\cite{palotas13fop} extended to include arbitrary tip orientations \cite{mandi13tiprot} emerges as a very promising tool to
explore the role of tip-orientations on the STM contrast of other surfaces.

\section{Conclusions}
\label{sec_conc}

In this work we studied the STM image contrast of the HOPG(0001) surface in the tunnelling regime as a function of the local
orientation of a set of tungsten tips. Employing a three-dimensional (3D) Wentzel-Kramers-Brillouin (WKB) tunnelling approach,
we demonstrated that the relative local orientation of the STM-tip apex with respect to the HOPG substrate can have a considerable
effect on the HOPG STM contrast. Depending on the STM tip-apex structure and composition, applied bias, and relative orientation
with respect to the substrate, substantially different effects, ranging from conservation to inversion of the STM contrast, were
observed. These results were rationalised in terms of the tip-rotation mediated contribution of tip-apex electronic states of
different orbital characters to the tunnelling current. For a sharp tungsten tip the HOPG contrast inversion between opposite bias
polarities was explained by the different weights of the tip orbital characters involved in the tunnelling that is due to the
asymmetry of the tip electronic structure with respect to its Fermi level.
We also compared the 3D-WKB and Bardeen STM simulation models with each other and with experiments in terms of
bias-voltage-dependent STM topography brightness correlations. We found quantitatively good agreement for particular tip models
and bias voltage ranges, and discussed the identified differences in view of the construction of the two tunnelling models.
In view of the experiments, we can also conclude that the two tunnelling methods perform at the same quantitative reliability.
Importantly for experimental STM analysis of HOPG, the simulations indicate that particular local tip-reconstructions with no
orientational change of the dominating $d_{3z^2-r^2}$ tip-apex orbital state affect only the secondary features of the HOPG STM
contrast, leaving the primary contrast unchanged, thus resulting in a stable tip. Such tip orientations are found to be
responsible for "striped" images observed in experiments. Conversely, tip-rotations leading to enhanced contributions from
$m\ne 0$ tip-apex electronic states can cause a triangular-hexagonal change in the primary contrast, indicating a likely tip
instability.

\section*{Acknowledgments}

The authors thank E. Inami, J. Kanasaki, and K. Tanimura at Osaka University for the experimental brightness data,
and A. L. Shluger for useful comments on the manuscript.
Financial support of the Magyary Foundation, EEA and Norway Grants, the Hungarian Scientific Research Fund project OTKA PD83353,
the Bolyai Research Grant of the Hungarian Academy of Sciences, and the New Sz\'echenyi Plan of Hungary
(Project ID: T\'AMOP-4.2.2.B-10/1--2010-0009) is gratefully acknowledged. G. T. is supported by EPSRC-UK (EP/I004483/1).
Usage of the computing facilities of the Wigner Research Centre for Physics, and the BME HPC Cluster is kindly acknowledged.

\appendix
\section*{Appendix: 3D-WKB tunnelling theory}
\setcounter{section}{1}
\label{sec_theory}

M\'andi {\it{et al.}} developed an orbital-dependent electron tunnelling model with arbitrary tip orientations
\cite{mandi13tiprot} for simulating scanning tunnelling microscopy (STM) measurements within the three-dimensional (3D)
Wentzel-Kramers-Brillouin (WKB) framework based on previous atom-superposition theories
\cite{tersoff85,palotas12orb,palotas11stm,palotas12sts,yang02,smith04,heinze06}. Here, we briefly describe this method used in the
paper for the highly oriented pyrolytic graphite, HOPG(0001) surface in combination with tungsten tips.
The model assumes that electrons tunnel through one tip apex atom, and individual transitions between the tip apex and a suitable
number of sample surface atoms, each described by the one-dimensional (1D) WKB approximation, are superimposed
\cite{palotas12orb,palotas11sts}. Since the 3D geometry of the tunnel junction is considered, the method is a
3D-WKB atom-superposition approach. The advantages, particularly computational efficiency, limitations, and the potential of the
3D-WKB method were discussed in Ref.\ \cite{palotas13fop}.

The electronic structure of the surface and the tip is included in the model by taking the atom-projected electron density of
states (PDOS) obtained by {\it{ab initio}} electronic structure calculations \cite{palotas11stm}. The orbital-decomposition of the
PDOS is necessary for the description of the orbital-dependent electron tunnelling \cite{palotas12orb}.
We denote the energy-dependent orbital-decomposed PDOS function of the $i$th sample surface atom with orbital symmetry $\sigma$
and the tip apex atom with orbital symmetry $\tau$ by $n_{S\sigma}^i(E)$ and $n_{T\tau}(E)$, respectively. In the present work
we consider $\sigma\in\{s,p_y,p_z,p_x\}$ atomic orbitals for the carbon atoms on the HOPG surface,
$\tau\in\{s,p_y,p_z,p_x,d_{xy},d_{yz},d_{3z^2-r^2},d_{xz},d_{x^2-y^2}\}$ orbitals for a blunt and sharp tungsten tip apex atom,
and $\tau\in\{s,p_y,p_z,p_x\}$ orbitals for a carbon apex atom on a sharp tungsten tip.
The total PDOS function is the sum of the orbital-decomposed contributions:
\begin{equation}
n_S^i(E)=\sum_{\sigma}n_{S\sigma}^i(E),
\end{equation}
\begin{equation}
n_T(E)=\sum_{\tau}n_{T\tau}(E).
\end{equation}
Note that a similar decomposition of the Green's functions was reported within the linear combination of atomic orbitals (LCAO)
framework in Ref.\ \cite{mingo96}.

Assuming elastic electron tunnelling at temperature $T=0$ K, the tunnelling current at the tip position $\mathbf{R}_{TIP}$
and bias voltage $V$ is given by the superposition of atomic contributions from the sample surface (sum over $i$) and the
superposition of transitions from all atomic orbital combinations between the sample and the tip (sum over $\sigma$ and $\tau$):
\begin{equation}
\label{Eq_current}
I\left(\mathbf{R}_{TIP},V\right)=\sum_i\sum_{\sigma,\tau}I_{\sigma\tau}^i\left(\mathbf{R}_{TIP},V\right).
\end{equation}
One particular current contribution can be calculated as an integral in an energy window corresponding to the bias voltage $V$ as
\begin{eqnarray}
I_{\sigma\tau}^i\left(\mathbf{R}_{TIP},V\right)&=&\varepsilon^2\frac{e^2}{h}\int_0^V T_{\sigma\tau}\left(E_F^S+eU,V,\mathbf{d}_i\right)\nonumber\\
&\times&n_{S\sigma}^i\left(E_F^S+eU\right)n_{T\tau}\left(E_F^T+eU-eV\right)dU.
\label{Eq_current_decomp}
\end{eqnarray}
Here, $e$ is the elementary charge, $h$ is the Planck constant, and $E_F^S$ and $E_F^T$ are the Fermi energies of the sample
surface and the tip, respectively. The $\varepsilon^{2}e^{2}/h$ factor ensures the correct dimension of the electric current.
The value of $\varepsilon$ has to be determined by comparing the simulation results with experiments, or with
calculations using standard methods, e.g., the Bardeen approach \cite{bardeen61}. In our simulations $\varepsilon=1$ eV was chosen
that gives comparable current values with those obtained by the Bardeen method \cite{palotas12orb} implemented in the BSKAN code
\cite{hofer03pssci,palotas05}. Note that the choice of $\varepsilon$ has no qualitative influence on the reported results.
The relative contribution of the $\sigma\leftrightarrow\tau$ orbital transition can be calculated as
\begin{equation}
\label{Eq_i_bg}
\tilde{I}_{\sigma\tau}\left(\mathbf{R}_{TIP},V\right)=\frac{\sum_i I_{\sigma\tau}^i\left(\mathbf{R}_{TIP},V\right)}{\sum_i\sum_{\sigma,\tau}I_{\sigma\tau}^i\left(\mathbf{R}_{TIP},V\right)}.
\end{equation}

In Eq.(\ref{Eq_current_decomp}), $T_{\sigma\tau}\left(E,V,\mathbf{d}_i\right)$ is the orbital-dependent tunnelling transmission
function, and it gives the probability of the electron tunnelling from the $\tau$ orbital of the tip apex atom to the $\sigma$
orbital of the $i$th surface atom, or vice versa, depending on the sign of the bias voltage. We use the convention of tip
$\rightarrow$ sample tunnelling at positive bias voltage ($V>0$), and sample $\rightarrow$ tip tunnelling at negative bias ($V<0$).
The transmission probability depends on the energy of the electron ($E$), the bias voltage ($V$), and the relative position of the
tip apex and the $i$th sample surface atom ($\mathbf{d}_i=\mathbf{R}_{TIP}-\mathbf{R}_i$). We consider the following form for the
transmission function \cite{mandi13tiprot}:
\begin{equation}
\label{Eq_Transmission}
T_{\sigma\tau}\left(E_F^S+eU,V,\mathbf{d}_i\right)=\exp\{-2\kappa(U,V)|\mathbf{d}_i|\}\chi_{\sigma}^2(\vartheta_i,\varphi_i)\chi_{\tau}^2(\vartheta_i',\varphi_i').
\end{equation}
Here, the exponential factor corresponds to an orbital-independent transmission, where all electron states are considered as
exponentially decaying spherical states \cite{tersoff83,tersoff85,heinze06}, and it depends on the distance between the tip apex
and the $i$th surface atom, $|\mathbf{d}_i|$, and on the vacuum decay,
\begin{equation}
\label{Eq_kappa}
\kappa(U,V)=\frac{1}{\hbar}\sqrt{2m\left(\frac{\phi_S+\phi_T+eV}{2}-eU\right)}.
\end{equation}
For using this $\kappa$ we assumed an effective rectangular potential barrier in the vacuum between the sample and the tip.
$\phi_S$ and $\phi_T$ are the electron work functions of the sample surface and the tip, respectively, $m$ is the electron's mass,
and $\hbar$ is the reduced Planck constant. The remaining factors of Eq.(\ref{Eq_Transmission}) are responsible for the
orbital dependence of the transmission. They modify the exponentially decaying part according to the real-space shape of the
electron orbitals involved in the tunnelling, i.e., the angular dependence of the electron densities of the atomic orbitals of
the surface and the tip is taken into account as the square of the real spherical harmonics
$\chi_{\sigma}(\vartheta_i,\varphi_i)$ and $\chi_{\tau}(\vartheta_i',\varphi_i')$, respectively. It is important to note that
the angles are given in the respective local coordinate system of the surface (without primes) and the tip apex
(denoted by primes). This distinction of the local coordinate systems is crucial to describe arbitrary tip orientations that
correspond to a rotation of the tip coordinate system by the set of Euler angles $(\vartheta_0,\varphi_0,\psi_0)$ with respect
to the surface coordinate system \cite{mandi13tiprot}. The polar and azimuthal angles given in both real spherical harmonics in
Eq.(\ref{Eq_Transmission}) correspond to the tunnelling direction, i.e., the line connecting the $i$th surface atom and the
tip apex atom, as viewed from their local coordinate systems, and they have to be determined for each surface atom from the
actual tip-sample geometry. A schematic view of an STM tip with rotated local coordinate system above the HOPG(0001) surface
is shown in Fig.\ \ref{Fig1}. For more details of the formalism, see Refs.\ \cite{palotas12orb,mandi13tiprot}.

\end{document}